\documentclass{KapProc} 
\usepackage{epsf}

\kluwerbib

 \normallatexbib 
\begin{document}
\newcommand{\hmp}{h^{-1}Mpc}      
\newcommand{\Ga}{\Gamma}     
\newcommand{\Om}{\Omega}     
\newcommand{\de}{\delta}     
\newcommand{\al}{\alpha}     
\newcommand{\si}{\sigma}     
\newcommand{\bx}{{\bf x}}     
\newcommand{\lam}{\lambda}     
\newcommand{\lan}{\langle}     
\newcommand{\ran}{\rangle}     
\newcommand{\La}{\Lambda}     
\newcommand{\bm}{\boldmath}     
\newcommand{\be}{\begin{equation}}     
\newcommand{\ee}{\end{equation}}     
\newcommand{\bea}{\begin{eqnarray}}     
\newcommand{\eea}{\end{eqnarray}}     
\newcommand{\ra}{\rightarrow}     
\newcommand{\bef}{\begin{figure}}     
\newcommand{\eef}{\end{figure}}     
\newcommand{\Mpc}{{\rm Mpc}}     
       
\newcommand{\veps}{\varepsilon}      
\def\spose#1{\hbox to 0pt{#1\hss}}      
\def\ltapprox{\mathrel{\spose{\lower 3pt\hbox{$\mathchar"218$}}      
 \raise 2.0pt\hbox{$\mathchar"13C$}}}      
\def\gtapprox{\mathrel{\spose{\lower 3pt\hbox{$\mathchar"218$}}      
 \raise 2.0pt\hbox{$\mathchar"13E$}}}      
\def\inapprox{\mathrel{\spose{\lower 3pt\hbox{$\mathchar"218$}}      
 \raise 2.0pt\hbox{$\mathchar"232$}}}      

\articletitle{Initial conditions, Discreteness and non-linear structure 
formation in cosmology}

\author{Francesco Sylos Labini}
\affil{$^1$Laboratoire de  Physique Th\'eorique,
 Universit\'e Paris XI, B\^atiment 211, F-91405   
Orsay, France}
\email{sylos@lyre.th.u-psud.fr}

\author{Thierry Baertschiger}
\affil{D\'ept.~de Physique Th\'eorique,  
            Universit\'e de Gen\`eve 24, Quai E. Ansermet,   
            CH-1211 Gen\`eve, Switzerland}
\email{thierry.baertschiger@physics.unige.ch}

\author{Andrea Gabrielli}
\affil{INFM Sezione Roma 1 \& 
Dip. di Fisica, Universita' ``La Sapienza'', 
P.le A. Moro 2 I-00185 Roma, Italy. }
\email{andrea@pil.phys.uniroma1.it}

\author{Michael Joyce}
\affil{Laboratoire de Physique Nucl\'eaire et de Hautes Energies,  
 Universit\'e de Paris VI, 4, Place Jussieu, 
Tour 33 -Rez de chaus\'ee, 75252 PARIS Cedex 05}
\email{joyce@lpnhep.in2p3.fr}

\begin{abstract}
In this lecture we address three different but related aspects 
of the initial continuous fluctuation field
in standard cosmological models. Firstly we discuss the properties of the  
so-called Harrison-Zeldovich like spectra. This power spectrum
is a fundamental feature of all  
current standard cosmological models.
In a simple classification of all 
stationary stochastic  processes into  three categories, we highlight  
with the name ``super-homogeneous'' the properties of the class  
to which models like this, with $P(0)=0$, belong. In statistical 
physics language they are well described as glass-like.
Secondly, 
the initial continuous density field with such small amplitude
correlated Gaussian fluctuations must be  discretised in order
to set up the initial particle distribution used in 
gravitational N-body simulations.  We discuss the main issues related 
to the effects of discretisation, particularly concerning
the effect of particle induced fluctuations on the statistical
properties of the initial conditions and 
on the dynamical
evolution of gravitational clustering.
\end{abstract}


In standard theories of structure formation in cosmology   
the matter density field in the early Universe 
is described as a homogeneous and isotropic   
distribution, with superimposed tiny fluctuations   
characterized by some particular correlation  
properties (e.g. \cite{pee93}). These fluctuations are believed   
to be the initial seeds from which, through a complex dynamical  
evolution,  galaxies and galaxy structures have emerged.  
In particular the initial fluctuations are taken to have 
Gaussian statistics and a spectrum which is, on large enough scales,
 the so-called {\it Harrison-Zeldovich} (hereafter H-Z)  
\cite{har,zel} 
or ``scale-invariant'' power spectrum (hereafter PS).   
Because fluctuations are Gaussian, the knowledge of the PS, or its  
Fourier conjugate, the real-space correlation function, gives 
a complete statistical description of the fluctuations. 
The H-Z type spectrum was first given a special importance 
in cosmology with arguments for its ``naturalness'' as an 
initial condition for fluctuations in the framework of  
the expanding universe cosmology, and it is in this context 
that the use of the term ``scale-invariant'' to designate it 
can be understood. 
It subsequently gained in importance  
with the advent of inflationary models in the eighties, and 
the demonstration that such models quite generically predict 
a spectrum of fluctuations of this type. Since the early nineties,  
when the COBE experiment \cite{cobe}   
measured for the first time fluctuations  
in the temperature in the Cosmic Microwave Background  
Radiation (hereafter CMBR) at large scales,  
and found results consistent with the predictions of models 
with a H-Z spectrum at such scales, the H-Z type spectra have  
become a central pillar of standard models of structure formation 
in the Universe. 

The H-Z spectrum arises in cosmology through a particular condition 
applied to perturbations of Friedman-Robertson-Walker (FRW) models, 
which  describe a homogeneous Universe in expansion
\footnote{Note that H-Z type spectra are the only ones
compatible with FRW metric on large scales, because 
they do not give divergent fluctuations of the gravitational
potential, as the purely Poisson distribution \cite{hz}}. 
This condition -  
commonly referred to in cosmology as ``scale invariance'' of the  
perturbations - gives rise to a spectrum (the  
``scale-invariant'' perturbation spectrum) with $P(k) \sim k$ 
at small $k$. 
All current standard cosmological models of structure formation in the 
Universe assume a spectrum exactly like this, or close to it, 
as initial condition (IC) for density 
perturbations in the Universe.  
In such models there is at any time only
one characteristic scale corresponding  
to the causal horizon, which increases with time, and below  
which causal physics can act to modify the spectrum.  
This causal physics depends, in general, on the details of the model,  
i.e. on the nature of its content in matter and radiation (or other 
forms of energy), until a characteristic time (the time when matter  
and radiation have comparable densities), after which purely  
gravitational evolution takes over.
In this lecture we firstly try  
to clarify the statistical properties in real space of these  
distributions, which have been almost completely overlooked in  
the literature on the subject (for further readings we refer
to \cite{hz,lebo}).

A related  context in which an understanding of the statistical  
properties in real space of the H-Z PS of the mass  
density field is important is represented by  
cosmological N-body simulations (hereafter NBS), the aim of which is to calculate the  
formation of structures under gravity in the Universe by a  
direct numerical calculation (see e.g. \cite{HE}). 
Since the time scale of evolution in these simulations is short 
compared to the dynamical time of the system (i.e. a particle  
moves a small distance relative to the size of the box 
representing a large volume of the Universe) the final  
configuration depends strongly on the IC at all but the smallest 
scales. Indeed a central idea is that from the  
final distribution - which should be closely related to 
the observed one of galaxies - one should be able to  
``reconstruct'' some important features of the IC, which can 
be related to other observations such as those of the CMBR.   

This leads to the problem of setting-up IC 
in standard cosmological NBS.
The problem concerns the setting-up of the {\it initial particle 
distribution}, as 
standard theories of galaxy formation predict
the properties of the {\it initial continuous density field}
with small amplitude
correlated Gaussian fluctuations. 
In general, most of the procedures
which for the discretisation of
a continuous field gives rise to Poisson noise
which dominates the dynamics at  small-scales.  
In order to avoid this 
and treat the dynamics 
due only
to  large scale (smooth) fluctuations, an ad-hoc method 
(lattice or glassy system plus
correlated displacements) has been 
introduced and used in  cosmological simulations.
We discuss the fact 
that such a  method gives rise to a particle distribution
which {\it does not have } any of the correlation properties
of the theoretical continuous density field \cite{thierry}. 
This is because 
discreteness effects, different from Poisson noise
but nevertheless very important, determine  particle fluctuations
at any scale so that the orginal conitinuos  fluctuations
become negligible.


The  third and last point  which we address in this lecture concerns the 
following question: 
which kind of fluctuations is the source of the gravitational
dynamics which gives rise to non-linear and power-law correlated
point structures ? 
In a particle distribution with large-scale
small-amplitude correlations there are 
two different sources of fluctuations which can drive 
the gravitational dynamics.
On one hand the intrinsic small-scale fluctuations
(which are inherent to any particle distribution) 
having   amplitude of order one at the smallest 
scales in the system (average distance 
between nearest-neighbors 
$\langle \Lambda_i \rangle$). On the other hand
there are,  on large scales,  some given 
small-amplitude (i.e. $ \delta \rho / \rho \approx 10^{-3}$) 
correlations
among density fluctuations.
We find  that 
in cosmological NBS 
small-scale particle fluctuations play an essential role in the 
development of power law correlations 
in the range $(0.03,3) \langle \Lambda_i \rangle$.
This 
implies that the  theoretical description 
should  focus on the discrete nature of the 
particle distribution. We refer the reader to
\cite{bjsl02,slbj02} for further readings.


\section{Real Space properties of H-Z like models}
 
Discussions of real space properties of the density fluctuations  
encountered  
in cosmology are puzzingly  
sparse in the literature on the subject. 
Peebles briefly notes (\cite{pee93} - see pg.523)  
that a  very particular 
characteristic of H-Z models is that  
``on large scales the fluctuations have to be anti-correlated   
to suppress the root mean square mass contrast on the scale   
of the Hubble length''.  Indeed, as we discuss below (and see
\cite{hz})  these models  
are characterized at large scales by a correlation function $\xi(r)$  
which has a negative power-law tail: detecting it would be 
the real space equivalent  of finding the turnover to  H-Z behavior  
to scales at which the PS  goes as $P(k) \sim k$.  
A basic question we try to answer is the following:  
What ``kind'' of two-point correlation function is the  
one corresponding to the H-Z behaviour in cosmological models?   
We compare it to some different statistical homogeneous 
and isotropic  systems:  
(i) Poisson-like distributions, (ii)  systems   
with a power-law correlation function found in critical phenomena  
\cite{ma84} and (iii) distributions characterized  
by long-range order (e.g. lattice or glass-like) \cite{hz,lebo}.  
Through this comparison we can classify H-Z models in the third 
category.  We introduce the term ``super-homogeneous''  
to refer to these kinds of distributions, as their primary  
characteristic is that mass fluctuations decay at large scales  
faster than in a completely uncorrelated (Poisson) system.   
For critical systems one has instead a decay of  
the (normalized) mass variance which is slower than Poisson.  
Formally the definition of this class of ``super-homogeneous'' distributions 
is given by the condition that the PS has $P(0)=0$, or  
equivalently in real space that the integral of the two point correlation  
function over all space is zero.

In the cosmological literature the latter property of 
cosmological models is often noted, but its meaning (as 
a strong {\it non-local}  
condition on a stochastic process) is not appreciated, 
or worse misunderstood as a trivial condition applying to any 
correlated system.
In the textbook of Padmanabhan\cite{padm}, 
for example, it is ``proved'' on pg. 171 that the integral over 
all space of the correlation function always vanishes
for any stationary stochastic process.  
The error is  
in an implicit assumption made that the number of particles  
in a large volume in a single realization converges exactly 
to the ensemble average. This is not true because, in general, 
extensive quantities such as particle number have fluctuations 
which are increasing functions of the volume (e.g. Poissonian, 
for which the integral of the correlation function over all space 
is not zero) even in homogeneous
systems. A slightly different, 
but common, kind of misunderstanding of the meaning of the  
vanishing of the integral over the correlation function is  
evidenced in the book by Kolb \& Turner \cite{kolbturner}. 
There it is affirmed (after its statement in Eq.(9.39)) to be  
``...just a statement  of mass conservation: if galaxies are clustered on   
small scales, then on large scale they must be ``anti-clustered''   
to conserve the total amount of mass (number of galaxies)''.  
The source of this misconception seems to be a confusion with the  
so-called ``integral constraint'' in data analysis  
(e.g. \cite{pee80,saslaw}), which imposes such a condition 
on the {\it estimator} of the correlation function in a {\it finite} sample, 
due to the fact that the (unknown) average number of points in such a volume  
is estimated by the (exactly known) number of points in the actual sample.  
Despite their apparent similarity, these are different conditions:  
the first (infinite volume) integral constraint provides non-trivial 
physical information about the intrinsic probabilistic nature   
of fluctuations, while the second is just an artifact of the  
boundary conditions which holds in a finite sample independently 
of the nature of the underlying correlations.  


\subsection{Statistical properties of statistically homogeneous 
and isotropic distributions}

We start by giving some basic definitions about the correlation
properties of statistically homogeneous and isotropic (SHI)
density fields. 
Inhomogeneities in cosmology are described using the general 
framework of stationary stochastic processes (hereafter SSP).  
Let us consider in general the description of a continuous or a discrete mass  
distribution $\rho(\vec{r})$ in terms of such a process. A stochastic  
process is completely characterized by its ``probability density functional''  
${\cal P}[\rho(\vec{r})]$ which gives the probability that  
the result of the stochastic process is the density field  
$\rho(\vec{r})$ (e.g. see Gaussian  
functional distributions \cite{Dobs}). For a discrete  
mass distribution  
the space (e.g. infinite three dimensional space) is divided 
into sufficiently small cells and the stochastic process  
consists in occupying or not any cell with a  
point-particle, and $\rho(\vec{r})$ can be  
written in general as:  
\be   
 \label{shi1a}   
 \rho(\vec{r})=\sum_{i=1}^{\infty} \delta(\vec{r} -\vec{r}_i)\,,          
 \ee      
where $\vec{r}_i$ is the position vector of the particle $i$   
of the distribution .

In a single realization of the mass distribution    
the existence of a well defined average density implies that      
\cite{gsl01,hz}  
\be     
\label{shi1}    
\lim_{R\rightarrow\infty}     
\frac{1}{\|C(R;\vec{x}_0)\|}\int_{C(R,\vec{x}_0)}d^3r\,   
\rho(\vec{r})=\rho_0>0    
\ee      
where $\|C(R,\vec{x}_0)\| \equiv 4\pi R^3/3$ is the volume of a sphere      
$C(R,\vec{x}_0)$ of radius $R$,  centered on the  
{\it arbitrary} point $\vec{x}_0$ of space  
\footnote{Because of the arbitrariness of the  
position of the center of the sphere,  
the average density is a  
one-point statistical property.}.    
When Eq.\ref{shi1} is valid one can then define \cite{gsl01,hz} a  
characteristic {\em homogeneity scale} as  
the scale $\lambda_0$ given by  
\be    
\label{shi2}    
\left| \frac{1}{C(R;\vec{x}_0)}   
\int_{C(R;\vec{x}_0)}d^3r\, \rho(\vec{r})-\rho_0 \right| < \rho_0    
\;\;\forall R>\lambda_0,\, \,\forall \vec{x}_0  
 \ee      
which depends on the nature of the fluctuations of the density in 
spheres (see \cite{hz} for more details).

\subsection{Real space correlation function}  
   
Let us analyze in further detail the auto-correlation properties of these   
systems. Due to the hypothesis of statistical homogeneity and isotropy,        
$\left<\rho(\vec{r_1})\rho(\vec{r_2})\right>$ depends only on       
$r_{12}=|\vec{r_1}-\vec{r_2}|$.          
The {\em reduced} two-point correlation function
$\tilde\xi(r)$      
is  defined by:     
\be     
\label{shi4}    
\left<\rho(\vec{r_1})    
\rho(\vec{r_2})\right> \equiv   
\rho_0^2\left[1+\tilde\xi(r_{12})\right]      
\ee
The correlation function $\tilde \xi(r)$ is one way
to measure the ``persistence of memory'' of spatial variations
in the mass density \cite{huang}.

Let us consider the paradigm of a stochastic homogeneous   
point-mass distribution: the {\it Poisson case}.  
For such a particle  distribution the reduced two-point correlation     
function Eq. (\ref{shi4})  can be written as (see \cite{gsl01})    
\be    
\label{p6}    
\tilde\xi(r) =   
\frac{\delta(\vec{r})}{\rho_0} \; \;\;(\mbox{i.e.}\;\,\xi(r)=0)\,.  
\ee    
The  previous relation is  a 
direct consequence of the fact      
that there is no correlation        
between different spatial points. That is, the reduced       
correlation functions $\tilde\xi$ has only the (diagonal) part
at $r=0$.       
The latter is present in the reduced correlation functions of any        
statistically homogeneous 
discrete distribution of particles with correlations.

In general \cite{landau,saslaw} for a  
SHI distribution of point-particles the reduced    
correlation function can be written as    
\be
\label{pc1}   
\tilde\xi(r)=\frac{\delta(\vec{r})}{\rho_0}+\xi(r)\,
 \ee
where $\xi$ is  the non-diagonal   
part which are meaningful only for $r>0$.     
In general $\xi(r)$ is a smooth function of $r$  
\cite{landau,gsl01}.

\subsection{Mass Variance}

Let us now  consider the amplitude of the mass fluctuations   
in a generic sphere of radius $R$ with respect to the average mass.   
First let  $M(R)=\int_{C(R)}\rho(\vec{r}) d^3r$  
be the mass (for a discrete distribution  
the number of particles) inside the sphere $C(R)$ of radius $R$  
(and then volume $\|C(R)\|=\frac{4\pi}{3}R^3$).   
The normalised mass variance is defined as     
\begin{equation}     
\sigma^2(R)=      
\frac{\langle M(R)^2 \rangle - \langle M(R) \rangle^2}   
{\langle M(R) \rangle^2}\,,     
\label{v1}     
\end{equation}     
where     
\begin{equation}     
\langle M(R) \rangle =    
 \int_{C(R)}d^3r \langle \rho(\vec{r}) \rangle=\rho_0\|C(R)\| \,,     
\label{v2}     
\end{equation}       
and    
\begin{equation}     
\langle M(R)^2 \rangle =    
\int_{C(R)} d^3r_1\int_{C(R)}d^3r_2 \langle  
\rho(\vec{r}_1)\rho(\vec{r}_2) \rangle\;.  
\label{v3}     
\end{equation}   
Note that there is no condition on the location of the center of the   
sphere, because of the assumed translational invariance of   
${\cal P}[\rho(\vec{r})]$.  
   
In the discrete Poisson case, using Eq. (\ref{p6}), we obtain that    
\be    
\label{v4}    
\sigma^2(R) = \frac{1}{\rho_0\|C(R)\|}\equiv \frac{1}{\left<M(R)\right>}  \;.  
\ee    
In general, for a   
SHI mass density field with     
correlations, substituting Eq.(\ref{shi4})   
in Eq.(\ref{v1}), we obtain  
\be    
\label{v6}    
\sigma^2(R) =  
\frac{1}{\|C(R)\|^2} \int_{C(R)} d^3r_1\int_{C(R)} d^3r_2   
\tilde\xi(|\vec{r_1} - \vec{r_2}|)  \;.  
\ee    
Using Eq.~(\ref{pc1}) in the discrete case we can write  
\be
\sigma^2(R) = \frac{1}{\rho_0\|C(R)\|}+     
\frac{1}{\|C(R)\|^2} \int_{C(R)} d^3r_1\int_{C(R)} d^3r_2   
\xi(|\vec{r_1} - \vec{r_2}|)  \;.  
\label{v6b} 
\ee
Note that the sign of the second term of Eq.(\ref{v6b}) is not uniquely   
determined. 
Equations (\ref{v6}) and (\ref{v6b}) make evident the relation between   
fluctuations in one-point properties (as in this case the number of   
points in a sphere centered on a random point in space)  
and two-point correlations.  

In the discrete case, to measure $\sigma^2(R)$ one has   
to take into account both terms in Eq.(\ref{v6b}),  
not only the second one.  
From Eq.(\ref{v6b}) the variance can, in general,  
be written as the sum of two contributions:    
\be    
\label{v9}    
\sigma^2(R) = \sigma^2_{Poi}(R) + \Xi(R)  \;,  
\ee    
where the first term $\sigma^2_{Poi}(R) \sim R^{-3}$    
represents the intrinsic Poisson noise of any stochastic     
particle distribution 
\footnote{Note that this term can give a contribution 
to the variance which dominates over that due to the 
intrinsic correlations.},   
and the second term $\Xi(r)$ (which, as noted above, does not  
have to be of a determined sign) is the additional contribution due to   
correlations (i.e. to $\xi(r)\ne 0$).  
Note that the first  term is specific of any {\it point} distribution,
in which fluctuations are never absent, and have large amplitude
(of order one) at small scale; in a continuous distribution, instead,
only the second term is present in both expressions
and fluctuations, which can be arbitrarily small at all scales,
are uniquely associated with correlations between different points.

To end this section note that 
Equation (\ref{shi1}) implies to the requirement that  
\be  
\label{v6a}  
\lim_{R \rightarrow \infty}  
\sigma^2(R) = 0 \;, 
\ee  
which is therefore a condition satisfied by any SHI  
distribution.  
Let us return to further discussion of Eqs.(\ref{v6}) and (\ref{v6b}).   
It is very important for our discussion to note that  
this condition (\ref{v6a}) which holds for any mass  
distribution generated by a SSP, is very different  
from the requirement   
\be  
\label{superhomo}  
\int d^3r   \tilde \xi(r) = 0  
\ee  
which is a much stronger special condition which holds for   
certain distributions - those to which below we will ascribe 
the name  ``super-homogeneous''.

\subsection{Power spectrum}

The PS $P(\vec{k})$ is the main statistical tool used to describe  
cosmological models. It is defined as 
\be 
P(\vec{k})=\lim_{V\rightarrow \infty}
\frac{|\delta_\rho(\vec{k})|^2}{V} = 
\left<|\delta_\rho(\vec{k})|^2\right> 
\label{ps-defn} 
\ee 
where 
\be
\delta_\rho(\vec{k}) = \int_v d^3 r e^{-i\vec{k}\vec{r}} 
\frac{\rho(r)-\rho_0}{\rho_0} \;,
\ee
limited to a volume $V$.
For a spatially stationary mass distribution $\rho(\vec{r})$ it is  
possible to demonstrate that it can be obtained by simply taking   
the FT of the correlation function  
$\tilde\xi(\vec{r})$ (up to a multiplicative constant) \cite{feller}:  
\be  
\label{lat7a}  
P(\vec{k}) = \frac{1}{(2\pi)^d} \int
d^dr \exp(- i \vec{k}\vec{r} ) \tilde  
\xi(\vec{r})\,.  
\ee  
Let us analyze the relation between the PS and the mass-variance 
in real space. 
We first rewrite Eqs.~(\ref{v1}-\ref{v3}), generalizing them to the case in   
which we calculate the mass variance in a topologically more complex volume  
${\cal V}$ of size $V$. To do this one introduces the window function  
$W_{\cal V}(\vec{r})$ defined as  
\be  
W_{\cal V}(\vec{r})=\left\{  
\begin{array}{ll}  
1\;\;\;\;\mbox{if}\;\vec{r}\in {\cal V}\\  
0\;\;\;\;\;\mbox{otherwise} \;.  
\end{array}  
\right.  
\label{window}  
\ee  
Therefore we can rewrite Eq.~(\ref{v2}) as  
\begin{equation}     
\left<M({\cal V})\right>= \int  W_{\cal V}(\vec{r})   
\left<\rho(\vec{r})\right> d^3 r\,.     
\label{mass2}     
\end{equation}  
and Eq.~(\ref{v3}) as  
\be  
\left<M^2({\cal V})\right>=\int\int 
d^3r_1 d^3r_2 W_{\cal V}(\vec{r}_1)  
W_{\cal V}(\vec{r}_2)  
\left<\rho(\vec{r}_1)\rho(\vec{r}_2)\right>\,,  
\label{v3bis}  
\ee  
where the integrals are over all space.  
The normalised variance is then given by 
\be  
\sigma^2({\cal V})=\frac{1}{V^2}\int
\int d^3r_1 d^3r_2   
W_{\cal V}(\vec{r}_1)W_{\cal V}(\vec{r}_2)\tilde\xi(\vec{r}_1-\vec{r}_2)\,.  
\label{sigma-w}  
\ee  
On taking the FT  one obtains  
\begin{equation}     
\sigma ^2({\cal V})=   
\frac{1}{(2\pi)^3} \int d^3 k P(\vec{k}) |\tilde{W}_{\cal V}(\vec{k})|^2     
\label{variance-ps}     
\end{equation}    
which is  explicitly positive, and $\tilde{W}_{\cal V}(\vec{k})$ is     
the FT of $W_{\cal V}(\vec{r})$, normalised by  the  
volume defined by the     
window function itself,         
\begin{equation}     
\tilde{W}_{\cal V}(\vec{k})=\frac{1}{V}     
\int d^3r e^{-i \vec{k}.\vec{r}} W_{\cal V}(\vec{r})     
\label{wfn-ft}     
\end{equation}     
with $V=\int_{\Omega} W_{\cal V}(\vec{r}) d^3 r$.

Consider now again the real sphere of radius  
$R$ for which the FT of the window function     
(normalised as defined) is     
\begin{equation}     
\tilde{W}_R(\vec{k})=\frac{3}{(kR)^3} 
\left( \sin kR - kR \cos kR \right) \;.    
\label{wfn-sphere}     
\end{equation}     
One then has, assuming statistical isotropy so that $P(\vec{k})=P(k)$, 
an expression for the variance in real spheres which is  
\begin{equation}     
\sigma ^2(R)=\frac{1}{2\pi^2} \int_0^\infty\! dk      
\frac{9}{(kR)^6}\left( \sin kR - kR \cos kR \right)^2 k^2 P(k)    \;. 
\label{sigma}     
\end{equation}     

One may show \cite{hz} 
that, for power-law spectra $P(k) \sim k^n$ 
(for small $k$, $n > -3$) the integral in (\ref{sigma}) has a  
very different behaviour for $n<1$ and $n \geq 1$ \cite{hz}.  
To summarize clearly: For a power-law  
$P(k) \sim k^n$ (with an appropriate cut-off around 
the wavenumber  $k_c$) the mass variance for real spheres with radius  
$R \gg k_c$ is given by  
 
\begin{enumerate} 
\item For $n<1$,  $\sigma^2(R) \sim 1/R^{3+n}$ and the dominant 
contribution comes from the PS modes at $k \sim R^{-1}$. 
\item For $n>1$, $\sigma^2(R) \sim 1/R^{4}$ and the dominant 
contribution comes from the PS modes at $k_c^{-1}$.   
\item For $n=1$, we have the limiting logarithmic divergence 
with $\sigma^2(R) \sim (\ln R)/R^{4}$. 
\end{enumerate}

In the cosmological literature 
\footnote{See, for example, the section entitled 
``Problems with filters'' in the book by Lucchin and Coles  
\cite{coles-lucchin}.} 
the divergences in the latter two  
cases are treated as a simple mathematical pathology due to the 
assumption of a perfect sphere (with a perfectly defined boundary). 
Replacing the real sphere with a smooth Gaussian filter  
$W_{\cal V}(\vec{r}) \sim e^{-r^2/R^2}$ these integrals are also 
cut-off at the scale $k \sim R^{-1}$ and one recovers a behaviour 
$\sigma^2 (R) \sim 1/R^{3+n}$ 
even for $n>1$. 
While of course this is valid mathematically 
it misses an important point, which is that this limiting behaviour  
of the variance (as $1/R^4$) has a very real physical meaning  
which has to do with the nature of systems with such a rapidly 
decaying PS. They correspond to extremely homogeneous 
systems (i.e. extremely ordered systems) in which the variance 
really is dominated by the small scale fluctuations. Let us 
explain this point further.

An intuitive understanding of this fact can be gained by considering 
discrete distributions. One would reason that any continuous distribution 
can be arbitrarily well approximated at large scales by an appropriate  
discretisation process, and that therefore the same result may hold of  
discrete distributions. 
In fact such a result has  
been proved several years ago \cite{beck}: 
In $d$-dimensions there exists no discrete distribution of points  
in which the variance in spheres decays faster than $1/R^{d+1}$.  
One can see roughly why this is so by considering the most ordered 
distribution of points one might think of: a simple cubic lattice. 
The variance in a sphere is given by averaging over spheres with 
center anywhere in the unit cell. As the sphere moves in the unit 
cell the variance, one would guess (correctly!),  
in the number of points is proportional to the difference in the volume 
of the spheres, which is proportional to the surface area of spheres,  
i.e. $\propto R^{d-1}$ in $d$-dimensions. Thus the normalised variance 
scales as $1/R^{d+1}$, a result proved in rigorously in  
\cite{kendall} (see also \cite{beck} for a more  
general discussion of the problem and the comments in \cite{hz}).   
They are distributions which are highly 
ordered (``glass-like'') in which the fluctuations in real space  
actually are at small scales (those at which the PS is 
cut-off). Because of this it is one of their characteristics,  
as we have seen, that {\it there is no direct relation between the  
PS  at scale $k$ and the physical variance in real space  
at the scale $R \sim k^{-1}$}.

\subsection{Real space classification of long range fluctuations}  
\label{classifications} 
  
We now return to a the discussion of the nature of correlations 
in systems with H-Z like power spectra, with the aim of elucidating 
their properties by comparison with systems described in  
statistical physics.  To this end, by following \cite{hz}, we
discuss a classification of all possible mass distributions in terms 
of the main features of the correlation function $\tilde \xi(r)$.  
Following from the above discussion 
concerning the behaviour of mass fluctuations,  
we define three distinct classes  
(for either the case of discrete particle distribution and  
of a continuous density field) \cite{hz}:    
\begin{enumerate}  
  
\item   
\label{1.} If    
\be  
\int  d^d r \; \; \tilde \xi(r) = const.>0   
\label{poi}  
\ee   
we can say that at large scale the system is   
{\em substantially Poissonian}. Indeed Eq.~(\ref{poi})  
implies that the PS goes to a constant non-zero 
value as $k$ goes to zero, and therefore that the 
large distance behavior of the mass fluctuations is  
\be  
\langle M^2(R) \rangle -\langle M(R)\rangle^2   
\sim R^d\sim \langle M(R) \rangle\,.  
\label{poi2}  
\ee  
We write here the unnormalized form of the variance, as the result 
that the variance of an extensive quantity such as the mass is  
proportional to the volume on which it is measured is the most 
intuitive way of characterizing a Poisson type behaviour.   
In this class is, for example, a system with a finite range  
correlation $\xi(r) \sim e^{-r/r_c}$. Beyond the scale $r_c$  
(the correlation length - see below 
for a discussion about this length)  
the system is uncorrelated and effectively 
Poissonian.  
 
\item \label{2.} If   
\be  
\int d^d r \; \; \tilde \xi(r) = + \infty   
\label{cri}  
\ee  
then we are in a case similar to a system at the critical point  
of a second order phase transition (e.g. the liquid-gas critical point).  
Such systems have a positive correlation function which is   
asymptotically a positive power law, with $\xi(r) \sim 1/r^{\gamma}$ and 
$\gamma < d$, corresponding to a PS $P(k) \sim k^{\gamma-d}$ 
as $k \rightarrow 0$. One then has at large scales the variance 
\be  
\langle M^2(R) \rangle -\langle M(R)\rangle^2   
\sim R^{\alpha}\;\; \mbox{with}\;d \le \alpha < 2d\;,  
\label{cri2}  
\ee  
or   $\langle M^2(R) \rangle -\langle M(R)\rangle^2\sim  
\langle M(R)\rangle^{\beta}$ with $\beta=\alpha/d>1$.  
This means that mass fluctuations are large (always overwhelming 
the Poisson fluctuations) and thus they are strongly 
and positively correlated 
at all scales\footnote{For 
example these properties near the critical point of  
the liquid-gas transition gives place to opalescence phenomena.}.  
It is in this context that the concept of self-similarity and 
scale-invariance has been introduced in statistical mechanics.  
These terms refer to the fact that in these systems the mass  
fluctuation field has well defined fractal properties \cite{slmp98}.

\item \label{3.} If  
\be  
\int d^d r \; \;\tilde  \xi(r) = 0  
\label{super}  
\ee  
then, as we have discussed, we have for the behaviour of the  
mass fluctuations 
\be  
\langle M^2(R) \rangle -\langle M(R)\rangle^2   
\sim R^{\alpha}\;\;\mbox{with}\;  
d-1<\alpha<d\,,    
\label{super2}  
\ee   
i.e. $ \langle M^2(R) \rangle -\langle M(R)\rangle^2  \sim   
\langle M(R)\rangle^{\beta}$ with $\beta=\alpha/d<1$, 
so that the mass fluctuations are always asymptotically 
smaller than in the uncorrelated Poisson case. This 
also corresponds to a strongly correlated, long-range ordered, 
system. We will refer to them with the term ``super-homogeneous'' 
to underline this feature that they are more homogeneous than a  
Poisson system. (Indeed, the Poisson particle distribution is  
considered as the paradigm  of a stochastic homogeneous mass  
distribution \cite{feller}). In the context of statistical mechanics  
they can be described as glass-like, as they have the properties 
of glasses, which are highly ordered compact systems. That can be said 
to be typically lattice-like, with a long-range ordered packing, but 
without the discrete symmetries of an exact lattice. Note again that,  
since $\tilde\xi(0)>0$ (a Dirac delta function in the discrete case)  
by definition, $\tilde\xi(r)$ must change sign with $r$ at least once. 
They are systems with finely balanced positive and negative correlation. 
\end{enumerate}  
 
The distinction between \ref{1.} and \ref{2.} is typical of the   
statistical physics of critical phenomena in order to distinguish  
a critical state (case \ref{2.}) from a non-critical state (case \ref{1.}).  
In this context the concept of correlation length is central.  
The correlation length is a measure of the distance up to which
one has spatial memory of the spatial variations in the mass 
density \cite{huang}.
There is no unique definition of this length scale, but from a  
phenomenological point of view it can be defined as the length  
scale up to which the effect of a small local perturbation in   
the system is felt. This is due to the {\em fluctuation-dissipation theorem}  
which links the response of the system to a local perturbation and the   
large scale behavior of the two-point correlation function  
(for the different precise definitions of the correlation length see  
for example \cite{ma84}). A simple definition is
\be 
r_{\rm{corr}}^2  
=\frac{\int r^2 d^d r \; \;  
|\tilde \xi(r)|}{\int  d^d r \; \; |\tilde \xi(r)|} \;. 
\label{corrlength} 
\ee 
In case \ref{1.} one can generally define a finite correlation length, 
while in case \ref{2.} it will generally diverge.  In particular  
in the case  $\xi(r)\sim \exp (-r/r_c)$, $r_c$ is indeed then the correlation 
length, while for a positive power-law $\xi(r) \sim 1/r^{\gamma}$ and 
$\gamma < d$ (case \ref{2.}) $r_{\rm{corr}} \rightarrow \infty$.

Case \ref{3.} is typical of ordered compact systems with small   
correlated  perturbations. One can meet this kind  
of correlation function for example   
in the statistical physics of liquids, glasses,  
phonons in lattices. The concept of correlation length in  
this context is less central, and the extension of its use to this class  
of systems is not particularly useful. Instead it is  
appropriate to classify the correlation properties of these 
systems directly through the integral of the correlation 
function as we have done. It is this behaviour of their correlations 
which distinguishes them from the other two cases, just as these 
cases are typically distinguished from one another by the 
value (finite or infinite) of their correlation length. 
Certainly, as we have noted, the use of the term ``correlation 
length'' in the cosmological literature, which is defined \cite{pee80} 
as a scale defining the amplitude of the correlation function, is 
in no way related to its use in statistical physics.

\subsection{CDM and HDM real-space correlation properties}

Let us firstly  consider a simple and instructive example: 
 a H-Z spectrum with a simple 
exponential cut-off:   
\be     
\label{exem1}     
P(k) = A\times k\times e^{-\frac{k}{k_c}}    \;, 
\ee     
where $A$ is the amplitude and $k_c^{-1}$ the cut-off scale.
The correlation function is given by 
\be     
\label{exem2}     
\tilde\xi(r) = \frac{A}{\pi^2} \frac{\left(\frac{3}{k_c^2}-r^2\right)}     
{\left(\frac{1}{k_c^2}+r^2\right)^3}    \;. 
\ee     
For $r < r_c \equiv k_c^{-1}$ we have  
$\tilde\xi(r) \simeq \frac{3A}{\pi}k_c^4>0$, changing  
at $r \sim r_c$ to an asymptotic behaviour $\xi(r) \sim -r^{-4}$. 
Note that the correlation does not oscillate, its only zero 
crossing being at scale $r= \sqrt{3} r_c$. Simply because of 
the condition $P(0)=0$, which implies that the integral of  
the correlation function must be zero, the  correlation 
function must change sign and in this case it only does 
so once and thus remains negative at large scales.

In the normalised mass variance $\sigma^2(R)$ shows  
a corresponding change in behaviour from 
being approximately constant at small scales  
$R<r_c$ to a $\ln R / R^{4}$ decay at large scales,  
as was shown  above. 
Note that, unlike for the  
variance in spheres, 
there is no limit to the rapidity of the decay of  
the correlation function (for the more general expression  
see \cite{pee80}). 
Despite the weakness of this correlation at large scales, 
however, the variance in spheres does not behave like  
that of a Poisson system, because of the  
balance between positive correlations  at small and negative  
at large scales imposed by the non-local condition $P(0)=0$.

In cosmological HDM  models the form of 
the PS is almost the same as we have just considered 
with an exponential cut-off \cite{padm}   
\be  
P(k) \sim k \exp(-k/k_c)^{3/2}  \;.
\ee  
A numerical integration verifies that the correlation function 
is essentially unchanged. 
 
For CDM  models, the class by far favored 
in the last few years, the form of the PS  at scales  
below turn-over from H-Z behaviour is considerably more  
complicated. In a linear analysis the PS  
of CDM matter density field decays below the turn-over with a 
power-law $\sim k^{-3/2}$ at large $k$  
until a smaller scale (larger $k$) at which  
it is cut-off with an exponential (in a manner similar  
to that in the HDM model). Numerical studies of these  
models designed to include the non-linear evolution  
bring further modifications, roughly increasing the 
exponent in the negative power law regime 
(see discussion in \cite{hz}).

In conclusion two simple real space characteristics today in the 
distribution of matter coming from the primordial H-Z PS 
are a {\it negative non-oscillating power-law tail in the two point  
correlation function $\xi(r) \sim -r^{-4}$, and a $(\ln R)/R^4$ decay  
in the variance of mass in spheres of radius $R$}.   
These are the peculiar distinctive feature of H-Z type 
spectra which should possibly be detected in real space by  
the new galaxy catalogs.

\subsection{Discussion}

The H-Z spectrum has the same behaviour characteristic of lattice-like 
order at large scales, while its small $k$ PS is $P(k) \sim k$  
instead of $\sim k^2$ \cite{hz,lebo},
typical of a lattice with completely random 
short range distortions. 
This spectrum corresponds to more power 
at large scales and it  can be  
associated with an appropriate correlated shuffling  
of a perfect lattice \cite{hz,lebo}. 
We thus say that the distribution described by the  
H-Z spectrum has a lattice-like or, more appropriately  
because of the isotropy, glass-like long range order. More 
specifically it can be described as a glass characterised by 
an opportune coherent long-range perturbative waves of 
distortions \cite{hz,lebo}.
These mass distributions  are so 
ordered at large scales that the mass variance 
at large scales really does come from small scales. 
The H-Z spectrum marks the transition to a 
pure lattice-like behaviour of the normalised 
variance in spheres $\sigma^2(R) \sim 1/R^4$, which 
has been shown to be the most rapid possible 
decay of this quantity for {\it any} stochastic distribution 
of points.

We now return to the use of the  term ``scale-invariance'' in  
cosmology: it refers to 
the fact that the variance of the mass (or equivalently gravitational 
potential) has an amplitude at the horizon scale which does not  
depend on time. The PS associated with this behaviour is 
that of a correlated system which is of the super-homogeneous type.  
This use of the term ``scale-invariance''  
therefore is not in any way analogous to its  
(original) use in statistical physics. In this context it is 
associated with a distinctly different class of distributions 
which have special properties with respect to scale  
transformations: typically critical systems, like a  
liquid-gas coexistence phase at the critical point,  
which have a well defined homogeneity scale and a  
reduced two-point  correlation function   
which decays as a non-integrable power law: $\xi(r) \sim r^{-\gamma}$  
with $0<\gamma<3$. In particular, is statistical physics,
the term does not have  
anything to do with the {\it amplitudes} of fluctuations
but with the persistence of the memory \cite{huang}.

We have highlighted the fact that all current cosmological models 
will share at large scales the characteristic behaviour in real 
space of the H-Z spectrum. Specifically we note primarily the 
very characteristic lattice-like behaviour of the variance  
in spheres  $\sigma^2(R) \sim R^{-4}$ (up to 
a small correction which is formally logarithmic for the case 
of exact H-Z), as well as the characteristic negative 
(non-oscillating) power-law tail in the two point correlation function 
$\xi(r) \sim -r^{-4}$. One would 
expect such behaviour to be seen in principle, if these models 
were correct, in the distribution of matter in the Universe 
at large scales, and in particular in the distribution of 
galaxies. So far such behaviour has not been observed.  
Rather the characteristic feature of galaxy clustering 
at small scales is that it shows fractal behaviour 
\cite{slmp98}, which corresponds 
to a very different kind of distribution than that  
described by CDM type models. A central (and 
much debated \cite{slmp98,joycesylos_ApJ2000}) 
 question is the determination of the scale  
marking the transition from this behaviour to homogeneity. 
%
In order to detect the correlations predicted by CDM  
in the distribution of galaxies, one should  
first find a clear crossover towards homogeneity 
i.e. a scale beyond which the average density becomes  
a well-defined (i.e. sample-independent)  
quantity \cite{joycesylos_ApJ2000,slmp98}. 
On much larger  scales galaxy structures 
should then present the super-homogeneous character  
of the H-Z type PS. Indeed this should be a critical  
test of the interpretation of measurements of CMBR  
in terms of the H-Z picture on large  
spatial scales \cite{cobe,boomerang}. 
Clearly the link between the observed fractal properties 
of the galaxy distribution and such super-homogeneous 
temperature fluctuations is a central problem for 
theoretical cosmology.  
Observationally a crucial 
question is the feasibility of measuring the transition 
between these regimes directly in galaxy distributions. 
With large forthcoming galaxy surveys it 
may be possible to do so, but this is a question which 
addresses exactly the statistics of these surveys and the 
exact nature of the signal in any given model.

 \section{Setting-up initial conditions in standard cosmological
simulations}

The purpose of cosmological NBS 
is to calculate the
{\it non-linear} 
growth of structures in the universe by following individual particles
trajectories under the action of their mutual gravity
\cite{HE}. These particles are not galaxies
but are meant to represent collisionless clouds of
elementary dark matter particles. 
In order to make them move, one must calculate the force
acting on each of them due to all the others. In general one may find
several algorithms  which speed up the $N^2$ sum necessary to
compute the force on each particle \cite{HE}. The force used
is not a pure $r^{-2}$ one:  instead one smoothes it at small $r$
by choosing for instance a force proportional to
$(r^2+\epsilon^2)^{-1}$ in order to avoid ``collisions'' between close
particles. This brings us to an important hypothesis sometimes made 
in cosmological NBS: 
with a softened force and a proper
choice of the softening parameter $\epsilon$, the evolution of the NBS
should be the same as the evolution of a continuous density field
(made of a huge number of  particles behaving like a
fluid) under its own gravity.
In a series of papers Melott \& collaborators \cite{me90,smss98,kms96} have
discussed the effects of discretisation in NBS, showing serious 
discrepancies in the dynamical evolution described by different
algorithms. In particular, and very importantly, they have questioned 
the capacity of high resolution NBS to describe 
correctly the evolution of a continuous density field. 
In our opinion this is still an open problem.
For example an important parameter is the ratio
$\epsilon/ \langle \Lambda \rangle$: in order
to simulate a self-gravitating fluid one would consider
that it should be 
larger than one, but this is not always the case.
For instance in the Virgo project
(nowadays the standard reference for simulations in the field) 
where $\epsilon = 0.036 Mpc/h$ \cite{jenkins98} 
and $\langle \Lambda \rangle  \approx 1 Mpc/h$ (see discussion
in \cite{thierry}) .

As already discussed, a  continuous and smooth density field 
with correlated density fluctuations 
is  given as IC.  However, if one wants to study the time evolution
of this field with NBS 
based on particle
dynamics, it is then necessary to discretise the field. This means
that one has to create a particle  distribution which 
is representative 
of the continuous density field and to control any finite size 
effect. In cosmology a certain procedure has been
used since twenty years: let us discuss it in more detail.

\subsection{The uniform background}
%
%

The standard ad-hoc procedure for setting up IC is described in 
\cite{efst85,white93,jenkins98}.  For the problem of
galaxy structure formation
 the IC generation splits into two parts.  The first is to
set up a ``uniform'' distribution of particles, which should 
represent
the unperturbed universe. The second is to impose density
fluctuations with the desired  characteristics.  
There are different procedures (e.g.  random sampling,
threshold sampling, etc.)  which can be chosen and they result in
different point distributions. Clearly one should have some physical
reasons to choose one or another since any procedure introduces some
discreteness effects, like Poisson noise, which  could 
play an important role in the non-linear dynamics of the system. 
For instance, in a Poisson
distribution
the dominant part of the gravitational force acting on an average
particle is due to its NN \cite{chandra43,gslp99}:
this is because statistical isotropy and the trivial
three-point correlation properties make the long-range
component of the force cancel.
If a
simulation is run from a pure Poisson IC
the intrinsic small scale fluctuations grow rapidly into
non-linear objects at small scales.
Instead,  in cosmological NBS, 
one would like to simulate 
a system where the main contribution 
to non-linear structure formation,
is due to the large scale
distribution of the
other particles and 
{\it not to local  NN interactions} \cite{white93}. 

To overcome the fact that a Poisson distribution 
leads to unwanted structure formation even without perturbations,
the most widely used solution has been to represent instead
the {\it unperturbed universe} by a {\it regular cubic grid of particles}
\cite{efst85,white93}. An infinite lattice, or a lattice
with periodic boundary conditions, is ``gravitationally
stable'' because  of symmetry.
However a lattice is a distribution with fluctuations at all scales
{\it and} non-trivial correlations.  
As discussed in the previous section,  the
unconditional variance in spheres of radius $R$ 
decays as $\sigma^2(R)\sim R^{-4}$.
The 
two-point
CF is such that $\xi(\vec{r}_1,\vec{r}_2) =
\xi(\vec{r}_1 -\vec{r}_2) \neq \xi(|\vec{r}_1 -\vec{r}_2|)$ because
it is not invariant for space rotation \cite{hz}: A lattice breaks space
isotropy.
Moreover,
the grid-like system has the disadvantage to introduce a strong
characteristic length on small scales - the grid spacing - and it
leads to strongly preferred directions on all scales.

\begin{figure}[tb]
\epsfxsize 8cm
\centerline{\epsfbox{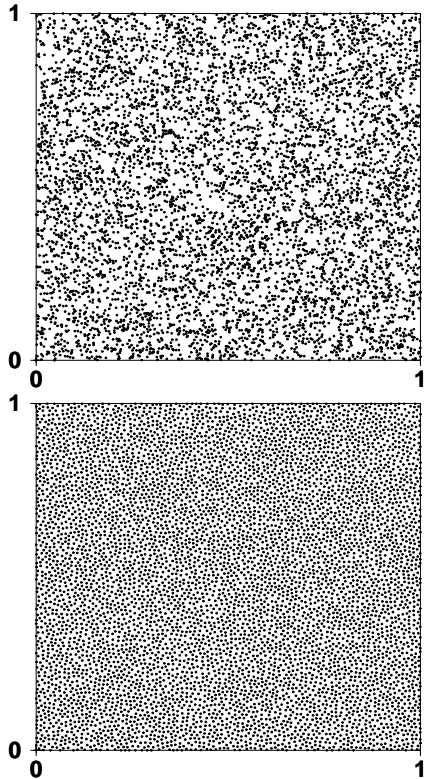}}
\caption{\label{fig2} In the top panel it is shown a Poisson 
distribution. while in the bottom one a glass-like 
distribution (see \cite{lebo} for details) }
\end{figure}

An alternative way to generate an ``uniform
background'' is by means of the following procedure.  
One starts from a Poisson distribution and then the N-body integrator is
used with a {\it repulsive} gravitational force in such a way that,
after the simulation is evolved for a sufficiently long time, the
particles settle down to a {\it glass-like configuration} in which
the force on each particle is very close to zero
\cite{white93}.  
The resulting distribution is very
isotropic but it is still characterized by 
long-range order of the
same kind  as in a lattice (see Fig.\ref{fig2}).  
As for the lattice the  distribution is
characterized by the  presence  
of an {\it excluded volume}:
two particles cannot lie at a distance smaller than a certain fixed
length scale \cite{hz}. In the lattice this scale is the grid space, for a
glass such a distance depends on the number of points one has
distributed in a given volume.  The fact that a lattice is ordered
is due to the existence of the deterministic small scale distance.
The unconditional variance scales as $\sigma^2(R) \sim R^{-\alpha}$
where $3< \alpha \le 4$, and it is again a strongly correlated
system \cite{hz}. 
Its two-point 
correlation function   depends on the detailed
procedures used to generate the glass distribution. 
As already mentioned, glass-like systems belong to a wide 
family of distributions for which the common feature is 
that $P(k) \sim k^a$ with $a>0$
\footnote{ To avoid any possible  
confusion for those somewhat familiar with these  
simulations, we note the description of the H-Z model 
as lattice-like or glass-like, has no direct relation to the 
use of lattices or glasses in setting up IC in current  
NBS. 
Their small $k$ behavior
is in general different as well as the real space 
correlation function \cite{hz}} for $k \rightarrow 0$  
and hence $P(0)=0$ \cite{hz,lebo}. However such behaviors in the PS 
do not imply directly that $\xi(r)$ has a negative  
power-law tail at large scales. In particular this is 
not true if the PS has a singularity for $P(0) \ne 0$,  
as happens in many systems. 

\subsection{Imposing correlated displacements...}


Given a ``suitably unperturbed'' particle distribution, any desired
{\it linear} fluctuation distribution can be in
principle generated using the {\it
Zeldovich approximation} \cite{efst85}.  Let us see 
how this method works for the ideal case of a continuous field.
Let the uniform density field be $\rho_0(\vec{r})=\rho_0$ and  
superimpose on it the stochastic displacement field $\vec{u}(\vec{r})$   
(the infinitesimal volume $dV$  
at $\vec{r}$ is displaced by $\vec{u}(\vec{r})$).  
Let us call $\rho(\vec{r})$ the resulting density field.  
We suppose that the stochastic field displacement is   
the realization of a stationary and isotropic  
stochastic process characterized by the  
probability density functional ${\cal P}[\vec{u}(\vec{r})]$.  
In this way ${\cal P}[\vec{u}(\vec{r})]$ defines also an ensemble   
of density fields $\rho(\vec{r})$ which is stationary and isotropic, 
with $\left<...\right>$ the ensemble average.  
  
By applying the mass conservation (i.e. the continuity equation) we find   
\be  
\frac{\rho(\vec{r})-\rho_0}{\rho_0}\simeq   
-\vec{\nabla}\cdot\vec{u}(\vec{r})\,.  
\label{dis}  
\ee  
If we call as usual $\xi(r)$ the reduced two-point correlation function  
of the density field we can write  
\be  
\xi(r)=\left<\vec{\nabla}\cdot\vec{u}(\vec{r})  
\vec{\nabla}\cdot\vec{u}(\vec{0})\right>\,.  
\label{xi-u}  
\ee  
Then, taking the FT of both sides of Eq.~(\ref{xi-u}),  
and making use of the statistical isotropy, we obtain  
\be  
P(k)\sim k^2 P_u(k)\,,  
\label{P-u}  
\ee  
where $P(k)$ is the usual PS of the mass density field  
and $P_u(k)$ is the PS of the displacement field.

As already discussed the pre-initial configuration 
can be a lattice or a glass-like system and hence 
one should consider the intrinsic fluctuations 
and correlation inherent to such a particle
distribution. This is not taken into account by the simple
approximation of Eq.\ref{P-u}. The standard procedure
neglects precisely these ``pre-initial'' perturbations.
In the continuous case, if the initial distribution
has a PS $P_i(k)$, what results from the infinitesimal
displacements is a PS
\be  
P(k)\sim k^2 P_u(k) + P_i(k)\,.
\label{P-u2}  
\ee  
The effect of the second term 
could be  important both 
for the determination of the statistical properties of IC,
and for the dynamical evolution, as we discuss below.
This is espcially true for the finite sample problem
(i.e. finite displacement and finite sample volume).
The analytical calculation in the discrete case is far more
complex than Eq.\ref{P-u2}.
In such a situation one has to
ensure that the correlations among density fluctuations implemented by
the displacement field are larger than the intrinsic fluctuations
of the original particle distribution, and that the large-scale fluctuations
dominate the non-linear small-scale clustering instead of 
nearest-particle interactions.
Only if one considers
very large scales, and/or a displacements field which introduces
correlations which are larger than the intrinsic one of the original
distribution, one can recover the PS as in Eq.\ref{P-u}.  
Otherwise
in a certain range of small enough scales,  
the point distribution is dominated
by discreteness effects, which in this context
can be seen as finite-size effects and which 
are important for what concerns the small-scale non-linear 
structure formation (see \cite{thierry,hz,lebo}).


In order to clarify the noise introduced by discretisation
and perturbations introduced with respect to the case
of the continuous field describe by Eq.\ref{P-u}
we have 
checked numerically the statistical properties of
the IC of some NBS. 
In \cite {thierry} we have analysed a CDM simulation with $\Omega_0=1$
\cite{jenkins98} as an example, but our result are generally valid
for any other particular model chosen, as they involve {\it the same method
for setting-up initial conditions}.  The result is that
there is a very bad between the real-space theoretical
correlation properties of the CDM continuous density 
field and the actual correlations of the particle
distribution used as IC (see Fig.2).

\begin{figure}[tb]
\epsfxsize 10cm
\centerline{\epsfbox{fig1.epsi}}
\caption{\label{fig1} Theoretical behaviour of the CDM correlation
function $\xi(r)$ (absolute value) 
for the IC of a NBS. The vertical
solid line correspond to the average distance between nearest neighbors:
at smaller scales the correlation function is oscillating due
to the imprint of the glass like properties. The horizontal dashed
line corresponds to the intrinsic lower limit (Poisson distribution)
in the case one uses all the $256^3$ to perform the statistical average.
One may see that even in this case that particle induced noise
fluctuations determine the behavior of the actual correlation function
for most of its range of scale. Finally the solid line is the determination
by \cite{thierry}.}
\end{figure}

\subsection{Discussion}

We have addressed the problem whether 
discreteness effects due to the imprint 
of the (correlated) fluctuations on the 
pre-initial point distribution 
are strong enough to dominate  
the fluctuations of the continuous distribution: 
the standard ad-hoc procedure
used to create NBS IC
{\it does not give rise to 
the desired CDM-like correlation of density fluctuations}.
This implies that  small-scale non-linear 
dynamical evolution
of the system is driven by fluctuations
which arise from the particular ad-hoc procedure 
used to discretise the field.
{\it In this context, these fluctuations 
can be seen as finite size effects, and 
are completely different from CDM-like fluctuations.}


\section{Discreteness and formation of non-linear power-law structures
in cosmological N-body simulations}


Non-linear structure formation represents a long-standing 
problem in cosmology. The central question concerns the 
organization of galaxies in 
the  highly irregular 
distribution characterized
by large  voids,  clusters and super-clusters 
which we  observe today in 
redshift surveys,
from some IC. 
There is a general agreement 
that the main feature of galaxy structures is their
fractal nature \cite{slmp98}, which is observed in the power-law
decay of the two-point correlation function. 
The  value of the fractal dimension 
and the possible detection of a well-defined crossover 
toward homogeneity are still matter of debate
\cite{slmp98,rees99}.
As already mentioned, according to standard 
theories of galaxy formation, IC of the mass distribution 
are represented by a continuous 
field with average density $\rho_0$ and with very small amplitude correlated
Gaussian fluctuations ($\delta \rho /\rho_0 \approx  10^{-5}$) 
\cite{padm,pee93}. 
NBS \footnote{
Note that there are basically two different families of N-body codes:
$P^3M$ and $PM$ \cite{white93}. We refer hereafter 
to the first kind of 
NBS, whose aim 
is to study  particle dynamics
with high small-scale resolution, as discussed below. 
Instead, $PM$ codes 
avoid close particles interactions \cite{me90,smss98,kms96}. 
} 
permit then to study numerically
the evolution of the discretised density field 
by particle dynamics \cite{white93,efst85,jenkins98}. 
 The final result of  
cosmological NBS, i.e. the  
particle distribution at the present time, should 
be 
similar to the observed galaxy structures
and it should  depend on the particular choice
of the statistical properties of the IC and 
of the various cosmological parameters 
\cite{padm,pee93}:
this is the crucial test for the 
different models, like Cold Dark Matter (hereafter CDM) 
and its variants.
%
%
%
%
NBS represent the primary tool 
which allows one to study  gravitational 
many-body dynamics in the non-linear regime (which corresponds
to the observed power-law correlated galaxy structures) 
and, possibly,  the transition
from linear fluid-like 
evolution to strongly non-linear clustering.
In fact, the understanding of the linear growth of 
structures, when the density field has small-amplitude
fluctuations, is based on linear 
perturbation theory of the equations of motion
of a fluid under the influence
of gravity in an expanding universe
\cite{padm}.
In this context, NBS are 
used as a guide for
any analytical approach to the 
weakly non-linear gravitational regime.

Given the fluctuations 
present in the system at the beginning
(particle fluctuations at small scales, small-amplitude
correlated fluctuations at large scales - which, we 
as discussed before, {\it are not} of CDM type)
which are the dominant ones for the generation
of non-linear structures with power-law correlations 
(as found in the cosmological NBS 
\cite{efst85,jenkins98}) ? 
In order to study this question, 
we characterize here the formation
of power-law correlated structures in cosmological 
NBS 
(based on particle dynamics
with a small initial velocity dispersion \cite{thierry})
by a detailed study of  real space statistical 
properties of the particle distribution during time evolution. 
Power-law structures seem to arise from  the (small-scale) 
particle fluctuations 
and  large-scale 
small-amplitude fluctuations appears to play little role
in this dynamics \cite{bjsl02,slbj02}.
Such a result implies a radical change of perspective
about the problem of structure formation, as
we discuss below.


We present here the analysis of a cosmological NBS performed
by the Virgo consortium \cite{jenkins98}, and we refer
the interested  reader to \cite{bjsl02,slbj02} for further 
discussions. This is a 
CDM model with $N=256^3$ particles and gravitational
force smoothing length $\epsilon = 0.036  \, Mpc/h$
\footnote{
With this smoothing the force is $53.6 \%$
of the true $1/r^2$ force at $\epsilon$ 
and more than $99 \%$ at  $2\epsilon$
\cite{jenkins98}.
}.  
The volume $V$ of the simulation is a cube of side
$L = 239.5 \, Mpc/h$. 
Each particle then represents  
a mass of $m_p= 2.27 \times 10^{11}$ solar masses, and 
 we find that the initial mean interparticle
separation $ \langle \Lambda_i \rangle  \approx 1 \, Mpc/h$.
The  NBS are all run from a
red-shift of $z=50$ until today ($z=0$), i.e. for a time
which is essentially the age of the Universe
\footnote{In these matter dominated cosmologies the 
time is given by $t=t_o/(1+z)^{3/2}$ where $t_o$ is the
age of the Universe today.}. This means that the time of 
evolution is essentially one dynamical time
$\tau_{dyn} \approx  1/\sqrt{G\rho}$, where the latter is 
simply the characteristic time scale associated with a 
mass density $\rho= N m_p/V $. A particle
typically moves by a distance of order the lattice spacing 
in this time. 

The first statistic
we consider is the conditional average density \cite{slmp98,hz} given
by 
$\Gamma(r) = \langle n(r) n(0) \rangle / \langle n \rangle$
where $n(r)$ is the microscopic number density. It is
simply the mean density of points at a distance $r$ from
an occupied point. It is plotted in Fig.\ref{fig1}
for a sequence of time slices, beginning from the 
initial distribution at $z=50$ until today. 
Also shown is the pre-initial
`glass' configuration.
\begin{figure}[tbp]  
\epsfxsize 12cm
\centerline{\epsfbox{FIG1nbs.epsi}}
\caption{Two-point conditional
density for the simulation considered.
The initial mean interparticle separation $\langle \Lambda_i \rangle$
and the softening length $\epsilon$ are shown, as well
together with the best power-law fit at the end of 
the simulation. In the insert panel the evolution
of the reduced correlation function $\xi(r)$ is shown.
\label{nbody} }
\end{figure} 
The behaviour of $\Gamma(r)$  for the glassy configuration 
and these IC at $z=50$ are very similar. This is because
the perturbations are very small in amplitude compared 
to the initial interparticle separation $\langle \Lambda_i \rangle$. The 
highly ordered lattice-like 
nature of the initial distribution is manifest:
there is an excluded volume around each point so that
$\Gamma(r)$ is negligible until very close to $\langle \Lambda_i \rangle$,
where it shows a small peak, with some oscillation about
the mean density evident corresponding to the long-range order 
\cite{hz}.
When the evolution under self-gravity starts the excluded volume 
feature is rapidly diluted, having almost completely disappeared by $z=3$ 
($t=0.125t_o$). This corresponds
to the development of power-law clustering at very small scales where
it was completely absent in the IC.

Let us consider what is driving this dynamics.
When points move 
towards their NN 
the 
contribution of the force acting on particle
due to its nearest neighbor grows as one over the distance squared
while the latter changes, supposing that the validity
of linear growth of fluctuations at larger scales,
in proportion to the scale factor, i.e. as $1/(1+z)$.
Thus at $z=5$ we would expect the latter to grow
by a factor of $10$ and thus the NN force to 
dominate below about $\langle \Lambda_i \rangle /5$.

Between $z=5$ and $z=3$ we see - in this range of scales 
completely dominated by NN interactions - the appearance of 
an approximately power-law behaviour in $\Gamma(r)$. At
$z=3$ the amplitude of $\Gamma(r)$ over most of the
range $r <\langle \Lambda_i \rangle $ is greater than 
the mean density so that this power law is also now
seen in the normalised correlation function 
$\xi(r) = \Gamma(r)/ \langle n \rangle -1$
shown in the inset of Fig.\ref{nbody}. At the next time slice, at 
$z=1$, the form of $\Gamma(r)$ up to approximately 
$0.3 \langle \Lambda_i \rangle$ is
almost precisely the same, being simply amplified by an
order of magnitude. At the same time the power law 
extends slightly further before flattening from 
$\Gamma \approx 3  \langle n \rangle$
to a smooth interpolation towards
its asymptotic value. The evolution in the remaining 
two time slices is well described over the range 
$\Gamma(r) > 3 \langle n \rangle$ as a 
simple amplification, and translation towards 
larger scales, of $\Gamma(r)$. At the final time
the power law (with exponent $\gamma \approx -1.7$) 
extends to $\approx 2\langle \Lambda_i \rangle$.

 In \cite{bottaccio02} a very similar behaviour 
is observed in the evolution of clustering of particles by
Newtonian forces, without expansion and with
simple Poissonian IC. 
The authors give a physical interpretation of this
clustering in terms of the exportation of the
initial ``granularity'' in the distribution to larger
scales through clustering. The self-similarity in
time of $\Gamma(r)$ is explained as due to a
coarse-graining performed by the dynamics: one 
supposes a NN dynamics between particle-like 
discrete masses, with the mass and physical 
scale changing as a function of time as 
the clustering evolves (particles forming
clusters, clusters forming clusters of clusters
etc.). This would appear to provide a good 
explanation for the behaviour observed 
here as well, but further theoretical 
work is clearly needed to establish this and
find a more quantitative  description.

The primary conclusion we draw from our real space 
analysis of the Virgo NBS is therefore that
the fluctuations at the smallest scales in
these NBS - i.e. those associated 
with the discreteness of the particles -
play a central role in the dynamics of clustering
in the non-linear regime. In particular the 
power-law type correlations appear to be built
up from the initial clustering at the smallest
scales. The nature of the clustering (in
particular the exponent of the power-law) seems 
to be independent of the IC, and 
its physical origin should  be explained 
through the dynamics of {\it discrete} self-gravitating 
systems. 
The fluid-like statistical description \cite{hz}  
and  equation of motions, which is the framework used to
describe a CDM universe, 
do not  consider  the non-analytical ``particle'' term  of noise
which is represented by NN interactions. The latter as we have
seen are strongly present in the NBS we have analysed and appear to
play a crucial role in the formation of the correlated
structures observed to emerge.
As already mentioned in the paradigmatic example
of stochastic (homogeneous and isotropic) point processes, the 
Poisson distribution, 
the gravitational force on an average point 
is due very predominantly to its NN \cite{chandra43}. 
Large-scale small-amplitude density fluctuations
do not give rise to a significant contribution
to the force acting a point, because of symmetry: i.e. 
large-scale isotropy \cite{chandra43}.

The dynamics we have described is 
essentially dependent on the gravitational forces at the
smallest resolved scale in the NBS, and 
the small smooth component of the force added to this
by the perturbations at larger scales appears to be
irrelevant in the non-linear regime. In cosmology this
supposed link between the IC and the ``predictions'' for 
structure formation at larger scales is very important,
as it is through it that one tries to constrain models
using observations of galaxy distributions. Our conclusions
completely change the perspective on this problem. Power-law 
correlations of this type are the most striking and
well established feature of such distributions 
\cite{pee93,slmp98}. The theoretical problem of
their origin therefore must deal with the apparently crucial
role in their formation of an intrinsically highly fluctuating
(and thus non-analytic) density field. In particular the
origin of the exponent in the correlations and
the dependence of the extent of such correlations 
on the discretisation (physical or numerical) needs
to be understood.


\begin{acknowledgments}
It is a pleasure to acknowledge B. Jancovici, J. Lebowitz 
and L. Pietronero for fruitful collaborations and discussions.
We thank M. Bottaccio, H. de Vega, R. Durrer, M. Montuori, 
and N. Sanchez for useful discussions and comments.
FSL acknowledge the support of a postdoctoral fellowship of the 
Swiss National Science Foundation.
\end{acknowledgments}

\end{document}